# Quantifying the form-flow-saltation dynamics of aeolian sand ripples

Joanna M. Nield[1]*, Matthew C. Baddock[2], Giles F.S. Wiggs[3], Jim Best[4], Kenneth T. Christensen[5], Pauline Delorme[6], Andrew Valdez[7], Nathaniel R. Bristow[8], Martin H.T. Hipondoka[9], Daniel P. Goss[1], Natasha S. Wallum[3], Philippe Claudin[10]

1 School of Geography and Environmental Science, University of Southampton, Southampton, SO17 1BJ, UK

2 Geography and Environment, Loughborough University, Loughborough, LE11 3TU, UK

3 School of Geography and the Environment, University of Oxford, Oxford, OX1 3QY, UK

4 Departments of Earth Science and Environmental Change, Geography & GIS, and Mechanical Science & Engineering, and Ven Te Chow Hydrosystems Laboratory, University of Illinois Urbana-Champaign, Urbana, IL 61801, USA

5 Department of Mechanical Engineering, University of Colorado Denver, Denver, CO, 80204, USA

6 Laboratoire de Géologie, Ecole Normale Supérieure, CNRS, PSL Research University, Paris, France

7 Great Sand Dunes National Park and Preserve, National Park Service, Alamosa, CO, USA

8 Astrin Biosciences, Inc., Saint Paul, MN, 55114, USA

9 Department of Environmental Science, University of Namibia, Private Bag 13301, Windhoek, Namibia

10 Physique et Mécanique des Milieux Hétérogènes, CNRS, ESPCI Paris, PSL Research University, Université Paris Cité, Sorbonne Université, Paris, France

Corresponding author: Joanna M. Nield (J.Nield@soton.ac.uk)

Key points:

1)      Ripple celerity exhibits a non-linear relationship to shear velocity under strong wind speeds.

2)      Ripple celerity and height respond more quickly to changes in wind speed than ripple wavelength and reorientation.

3)      Aerodynamic roughness is influenced by saltation and ripple height, particularly under stronger winds transitioning to a collisional regime.

Abstract

Ripples are the most fundamental and ubiquitous of aeolian bedforms formed on sandy surfaces, but their small size and fast response times make them inherently difficult to measure. However, these attributes also make ripples excellent flow indicators, and they have been used extensively in planetary locations for this purpose.  Here, we use terrestrial laser scanning to measure ripple morphometry and celerity coincidently, as well as saltation height above rippled surfaces.  We find that although ripple height and wavelength respond linearly to increased shear velocity, under strong winds ripple celerity exhibits a non-linear increase.  This relationship at high wind speeds is also reflected in the response of aerodynamic roughness and saltation dynamics, with a greater maximum in saltation height present over ripple lee slopes. Importantly, when using ripple patterns as indicators of flow conditions, celerity or height should be used in preference to wavelength as their dynamics respond faster to changing wind speed. In planetary and stratigraphic settings where measuring celerity is not possible, wavelength should be considered as indicative of consistent wind conditions, rather than the full range of sand transporting windspeeds.

Plain Language Summary

When wind blows over a sandy surface, it typically shapes the sediment into small wave-like patterns known as ripples, which are typically a few millimetres high and tens of centimetres long.  These ripples exist on sandy surfaces in coastal and desert areas on Earth, as well as other planetary bodies.  The height and spacing of ripples changes with the velocity of the wind, with both the size and the speed at which the ripples move typically increasing at greater wind speeds.  If we can relate the size and speed of ripple movement to wind conditions, then we can potentially infer the winds that formed the ripples through measuring the ripples, rather than measuring the wind.  This is particularly useful in environments where wind measurements are difficult or impossible.  Here, we find that the height and movement (or migration rate) of ripples respond more quickly than the spacing between ripples to increases in wind speed, or the orientation of the ripples when the wind direction changes. Consideration of the differing responses of ripple characteristics to changing wind conditions improves our ability to use their shape and size on Earth and other planets to quantify ripple dynamics and their formative winds.

1. Introduction

Wind-ripples (also referred to as splash or impact ripples) typically form on sandy surfaces where saltation occurs (Bagnold, 1941; Seppälä and Lindé, 1978). Although ripples are the most common and responsive of aeolian bedforms, there has been little research linking their morphological and migratory dynamics to flow or transport drivers (Andreotti et al., 2006; Sherman et al., 2019a). Both modelling and experimental data have elucidated the critical role of grain-bed impacts in driving these dynamics (Anderson, 1987; 1990; Andreotti et al., 2006; Duran et al., 2014; Lester et al., 2025; Yizhaq et al., 2004; Yizhaq et al., 2024), or the role of mid-air particle collisions (Huo et al., 2024) that increase in frequency with increased wind shear velocity and shift the regime of sediment transport to one that is collision-dominated (Pähtz and Durán, 2020; 2023; Ralaiarisoa et al., 2020). While it has been established that ripple celerity (or migration rate) increases with faster flow (Andreotti et al., 2006; Duran et al., 2014; Sharp, 1963; Sherman et al., 2019a; Uphues et al., 2022), there is disparity between field and wind tunnel measurements as to the exact rate at which celerity increases (Sherman et al., 2019a).

Studies on the behaviour of aeolian ripple morphology typically only consider planform ripple shape and ripple wavelength, or measure a single cross-profile transect line, rather than the full 3D ripple form. This limitation is partly due to previous methods being more conducive to analysis in 1 or 2 D, for example manual measurements such as the shadow method (Werner et al., 1986; Zimbelman et al., 2012), time-lapse photography (Lorenz, 2011; Lorenz and Valdez, 2011; Yizhaq et al., 2008) and laser sheet approaches (Sherman et al., 2019a). While model (Anderson, 1987; Duran et al., 2014) and wind tunnel (Gordon and McKenna Neuman, 2011) results suggest asymmetry in form-flow-transport interactions between the ripple stoss and lee slopes, with a greater concentration of particle collisions and ejections on the ripple stoss, evidence of this behaviour in a field context is lacking. Yet, as these universal bedforms are inherently responsive to flow, the ability to reconstruct wind and transport conditions remotely based solely on imagery is enticing, and this would have applications in instances where ripples are used as flow indicators, such as on Mars (Ewing et al., 2017; Hood et al., 2021; Lapôtre et al., 2021; Lapotre et al., 2016; Liu and Zimbelman, 2015; Roback et al., 2022; Rubanenko et al., 2022; Silvestro et al., 2010; Sullivan et al., 2020; Vaz et al., 2023).

Work by Owen (1964) has highlighted the role of the saltation cloud in driving changes in aerodynamic roughness ($z_0$), in contrast to estimates of grain-scale roughness (Nikuradse, 1933). However, ripples are roughness elements and surface roughness also influences aerodynamic roughness (Duran et al., 2019; Field and Pelletier, 2018; Nield et al., 2013; Pelletier and Field, 2016; Sherman and Farrell, 2008). While $z_o$ has been studied extensively in both field (Furtak-Cole et al., 2022; Gillies et al., 2007; King et al., 2006; Lancaster and Baas, 1998; Marticorena and Bergametti, 1995; Raupach, 1992; Raupach et al., 1993; Shao et al., 2015; Wolfe and Nickling, 1996) and wind tunnel experiments (Alvarez et al., 2025; Brown et al., 2008; Cheng et al., 2007; Gillies et al., 2017; King et al., 2008), it is typically parameterised for discrete roughness elements, such as vegetation, rather than continuous and complex rough surfaces such as aeolian ripples.

Terrestrial Laser Scanning (TLS) has revolutionised how we can characterise surface roughness, and several studies have examined the influence of high-resolution surface characterisations both with (Field and Pelletier, 2018) and without saltation (Nield et al., 2013; Pelletier and Field, 2016). These studies found that without saltation, aerodynamic roughness is most strongly influenced by surface roughness element height, whereas with saltation, the characteristics of the saltation cloud itself are more important in determining $z_0$. In field environments, it is difficult to determine whether the physical height of a ripple, or the thickness of the saltation layer, has a greater impact on the aerodynamic roughness. Experimental data concerning the multiple scales of aerodynamic roughness on Earth is thus needed urgently (Cooke et al., 2025; Jia et al., 2023) as accurate predictions of aerodynamic roughness are crucial for modelling shear velocity and sediment flux (Farrell and Sherman, 2006).

Here, we employ TLS to quantify, for the first time, both ripple morphology and celerity, as well as the concurrent saltation dynamics, above ripples under varying wind conditions. We elucidate the role of saltation layer depth in ripple adjustment and identify important form-flow-transport interactions close to, and at the same vertical scale as, an erodible sandy surface.

## 2. Study site and methods

Four experiments were undertaken on a flat surface within the dry, sand-covered bed of Medano Creek, Great Sand Dunes (GSD) National Park and Preserve, Colorado, USA in 2022 and 2023 (Table 1 and Figure 1) (Nield et al., 2025a; Nield et al., 2023a). Rippled surfaces, and above surface saltation, were measured using Leica TLS Scanstations (280 scans) over an approximately 1 $m^2$ area immediately upwind of a Campbell Scientific CSAT3 3D sonic anemometer that measured wind speed and direction at 0.24 m above the surface. The uniform ripple-scale local surface roughness and an unperturbed upwind fetch of 100s m, resulted in the anemometer measurements being from within the subregion of the boundary layer where shear stress is constant. Under the conditions of a fully developed boundary layer, small-scale changes in roughness imposed by the active saltation cloud or ripple height are assumed to not meaningfully alter the location of the anemometer in relation to the constant depth shear layer. Saltation flux measurements were recorded immediately downwind of each scanned square scanned area using: i) a Sensit-piezoelectric counter (Van Pelt et al., 2009) that was positioned so that the sensor base was flush with the surface and the sensor top extended to a height of 0.014 m above the surface, and ii) Wenglor YH03PCT8 optical gate sensors (Hugenholtz and Barchyn, 2011) at heights of 0.02 m and 0.05 m (Figure 1a). Measurement duration was 3 minutes or 1.5 minutes for weaker and stronger winds respectively. Saltation heights were calculated as mean maximum heights detected by the TLS following the methods of Nield and Wiggs (2011), and differ from heights derived from exponential fits of flux curves (e.g. Ho et al., 2011; Martin and Kok, 2017). Further details on data processing methods are given in the Supplementary Information and are similar to those presented in Delorme et al. (2023).

**Table 1**: Details of the TLS measurements for each experimental set-up, Great Sand Dunes National Park and Preserve, USA

| Set-up | Date | Leica Instruments | Number of scans | TLS head height (m) | Distance to ripple patch from TLS head (m) | Mean $u_*$ (m/s) | Standard deviation of $u_*$ (m/s) | Initial wind direction (°) |
|---|---|---|---|---|---|---|---|---|
| 1 | 5th April 2022 | P20 | 62 | 1.86 | 7.7 | 0.38 | 0.078 | 253 |
|   |                | P50 | 59 | 1.87 | 7.9 |      |       |     |
| 2 | 30th March 2023 | P50 | 48 | 1.85 | 8.2 | 0.54 | 0.063 | 53 |
| 3 | 3rd April 2023 | P50 | 49 | 1.81 | 7.9 | 0.53 | 0.054 | 220 |
| 4 | 4th April 2023 | P50 | 62 | 2.03 | 7.5 | 0.41 | 0.075 | 245 |

Each approximately 1 x 1 m rippled surface was divided into 0.002 m transects, parallel to the wind direction. Ripple heights were calculated for each transect using the zero-upcrossing method (Davis et al., 2004; Goda, 2000; Martin and Jerolmack, 2013). Ripple wavelengths and celerities were calculated for each transect and transect pair respectively using the Matlab cross covariance function. Checks were performed to ensure that the distance that the ripples migrated between scans was less than one bedform wavelength. Bulk relationships of ripple celerity, height and wavelength were calculated using data that had a variation in wind direction of less than 15° from the initial direction of ripple migration.

Previous studies have estimated sediment transport via ripple dynamics through ripple celerity, $c_r$, and either a ripple wavelength, $l_r$ -derived transport flux, $q_l \propto c_r l_r$, (Duran et al., 2014) or a ripple height, $h_r$ -derived transport flux, $q_h \propto c_r h_r$, (Jerolmack et al., 2006). We compare our field measurements of sediment flux to relative ripple-derived flux to examine which ripple metric is a better indicator of flux.

We identified ripple crests using the Matlab edge detection algorithm over the gradient of the surface using the Canny algorithm, followed by deblurring and selection of lines that were above a mean height over the surrounding 0.5 x 0.5 m area, similar to methods used to identify dune crests by Daynac et al., (2024) and Hugenholtz and Barchyn (2010). Crests were then classified as mature if their length was >90% of the measured surface width (i.e. 0.9 m). Lags in ripple responses to changes in wind conditions were calculated using cross-covariance.

Additional measurements (104 scans) were collected on flat crestal areas of barchan and dome dunes in the Huab Valley, Skeleton Coast, Namibia using the same TLS methodology for ripple morphometry, and either 3D sonic or cup anemometers for mean wind speed measurements only and $z_o$ from the literature to estimate $u_*$, (Table 2; Supplementary Methods) (Nield et al., 2023b). Celerity and height data from Oceano Dunes (Sherman et al., 2019b), reported in Sherman et al. (2019a), were also used as a comparator dataset as these were collected using a methodology similar to our own, including the use of a 3D sonic anemometer to calculate shear velocity ($u_*$) and a single laser to measure topography. While Sherman et al. (2019b) also report wavelength data, this did not use the same laser technique employed herein, and thus this part of their dataset is not directly comparable to the present paper. Both the Huab and Oceano data supplement our main GSD dataset as they were collected at lower shear velocities. Where multiple field site data are used, we have normalised the data to account for grain sizes.

**Table 2**: Details of the TLS measurements for additional measurements in the Huab Valley, Namibia

| Dune Type | Date | Leica Instruments | Number of scans | TLS head height (m) | Distance to ripple patch from TLS head (m) | Mean $u_*$ (m/s) | Standard deviation of $u_*$ (m/s) | Anemometer type | Grain size (µm) |
|---|---|---|---|---|---|---|---|---|---|
| Barchan | 6th September 2014 | P20 | 25 | 1.96 | 13 | 0.29 | 0.013 | 3D sonic (0.5 m above surface) | 341 |
| Dome | 22nd September 2016 | P20 | 61 | 1.78 | 5 | 0.32 | 0.021 | Cup (0.24 m above surface) | 402 |
| Dome | 8th September 2018 | P20 | 18 | 1.71 | 7 | 0.26 | 0.011 | Cup (0.24 m above surface) | 402 |

Field quantification of aerodynamic roughness, $z_o$, was undertaken using the Law-of-the-Wall from measurements of Reynolds stress derived $u_*$, and the mean velocity measured at a height of 0.24 m (van Boxel et al., 2004). These measurements were compared to standard empirical relationships between $z_o$ and $u_*$ using the Bagnold roughness law (e.g. Valance et al., 2015; Duran et al, 2011; Creyssels et al., 2009), Equation 1 (where $z_f$ is the focus height, κ is the von Kármán constant and $u_f$ is the wind velocity at $z_f$ ), and a Charnock type model (e.g. Sherman and Farrell, 2008), Equation 2, where C is the Charnock constant and g is gravitational acceleration.

$$z_o = z_f exp\left(\frac{-\kappa u_f}{u_*}\right) \quad (1)$$

$$z_o = \frac{Cu_*^2}{g} \quad (2)$$

The relationship between $z_o$ and the surface elevation profile was also characterised in the absence of saltation using the empirical model of Nield et al. (2013). In this case, Jia et al. (2023) found that ripple scale roughness in the absence of saltation should be within the transition between a smooth (Nikuradse, 1933) and rough (Flack and Schultz, 2010) roughness regime, while the presence of saltation moves the system to a rough regime.

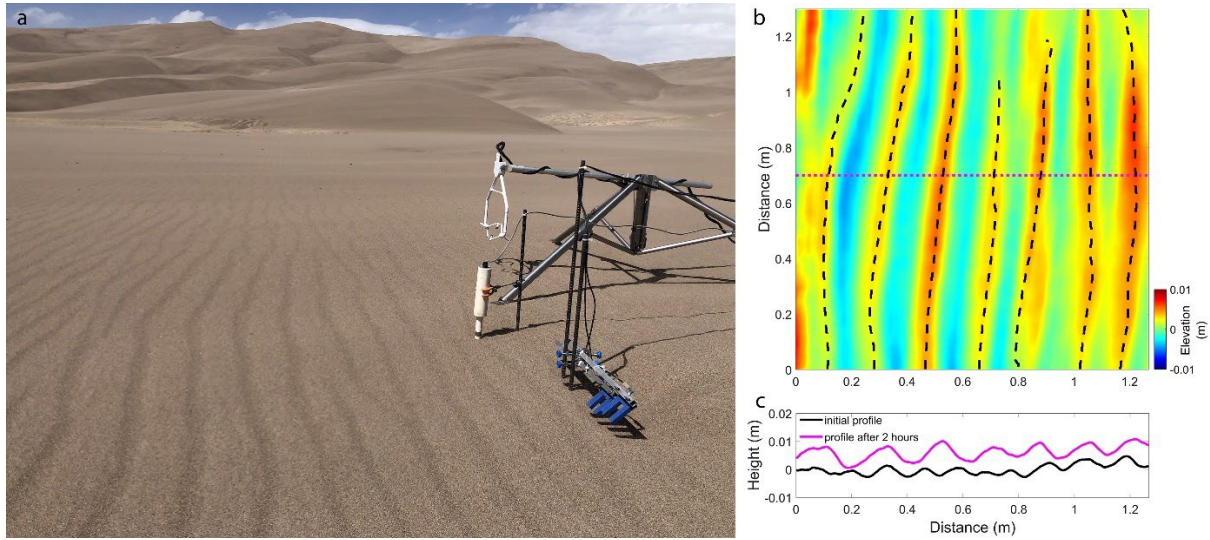

Figure 1: a) Field site set-up at Great Sand Dunes National Park and Preserve on 3rd April 2023. Wind direction from left to right. b) Example of rippled surface measured on 30th March 2023, with black dashed lines being the identified crest orientations and the magenta dotted line showing the location of the ripple cross sections (c).

3. Results and Discussion

3.1. Shear velocity drives ripple height, wavelength and celerity

In general, we find that ripple celerity, wavelength and height all increase with increasing shear velocity (Figure 2). The mean ripple index ($l_r/h_r$) within our time averaged datasets remains constant (mean = 41, standard deviation = 2.4). Similar to previous studies that measured ripple dynamics at low values of shear velocity (<2.5 $u_{*t}$) (Duran et al., 2014; Sherman et al., 2019a), we find a linear relationship between $u_*$ and both ripple height and wavelength across a wider range of $u_*$ (1<$u_{*t}$<3.5, Figure 2b-c). However, using this greater range of $u_*$, which was not accounted for in the previous studies of Duran et al. (2014) and Sherman et al., (2019a), our data also identify a non-linear relationship between $u_*$ and ripple celerity (Figure 2a). The reasons for this are not clear. However, the flattening of ripples under strong winds has been noted by other researchers (Bagnold, 1941; Sharp, 1963) and it is possible that under stronger winds, the increased height of the ripple relative to the mean surface height enables an increased shear velocity to be experienced on the ripple crest, acting to increase local erosion and ripple celerity. Alternatively, under strong winds, the system might be transitional from a saltation- to a collision-characterised transport regime, where saltating grains collide dominantly with other grains above, rather than on, the surface, thereby travelling higher and faster and increasing the sediment flux (Pähtz and Durán, 2020; Ralaiarisoa et al., 2020). It is unclear if the threshold to a collisional regime measured in the wind tunnel is similar to a field environment, and our shear velocity values are lower than those reported by Ralaiarisoa et al. (2020). However, the possible transition towards the collision regime is supported by our celerity fit (Figure 2a) that switches from linear to cubic for higher $u_*$. This is in agreement with the quartic sediment flux relationship reported in Pähtz and Durán (2020) where sediment flux is equivalent to the product of celerity and ripple height or wavelength (Duran et al., 2014; Jerolmack et al., 2006), with our height and wavelength relationships remaining linear (Figure 2b-c). The dynamics of mid-air collisions in saltation would act to accelerate ripple celerity due to a greater relative flux acting on the ripple surface. More work is required to investigate whether these, or other factors, might be responsible for the cubic response of ripple celerity with stronger winds.

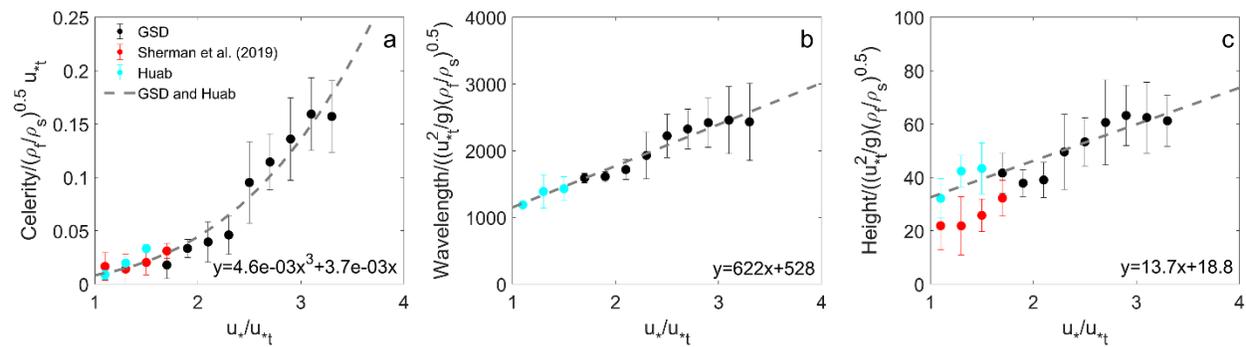

Figure 2: Normalised relationships between ripple a) celerity, b) wavelength and c) height and shear velocity, with comparisons to other datasets, where g is gravity, $\rho_f$ is fluid density and $\rho_s$ is sediment density. Error bars indicate standard deviation within each normalised shear velocity bin. The $R^2$ values for dashed line fits are 0.95, 0.96 and 0.82 for ripple celerity, wavelength and height respectively.

3.2. Ripple dynamics, flux indicators and lagged response to changes in wind speed and direction

When comparing the response of ripple celerity, height and wavelength to changes in shear velocity, we find that celerity is the first geometric attribute to respond and exhibited a mean lag of 2.3 ± 1.1 minutes (Table 3). These time scales of several minutes make physical sense as they are typically ca. 10-20 times larger than the initial growth times estimated by linear stability analysis in the numerical simulations of Duran et al (2014). Ripple height is the second fastest attribute to respond with a mean lag of 5.6 ± 1.7 minutes, while ripple wavelength was the slowest to respond to changes in wind speed (8.3 ± 1.4 minutes). In all datasets (both from GSD and the Huab), wavelength responded more slowly than either one or both celerity and height, irrespective of the magnitude of shear velocity or the variation in shear velocity driving the response (Table 3). This suggests that while ripple celerity is the most responsive indicator of shear velocity, ripple height is a better indicator of shear velocity than wavelength if wind conditions are

author accepted manuscript
Nield et al. (2025) Journal of Geophysical Research: Earth Surface, 130, e2025JF008616
http://dx.doi.org/10.1029/2025JF008616 - available open access from publisherfluctuating. Caution is thus advised when using wavelength as an indicator of shear velocity unless conditions have been consistent for ≥8 minutes, depending on the wind strength (Table 3). Furthermore, ripple morphometry might not represent the most recent wind conditions if the wind speed was reducing and there was insufficient time for the ripples to adjust.

Previous research has suggested that some function of ripple celerity, and either height (Jerolmack et al., 2006) or wavelength (Duran et al., 2014), could be used as a proxy to measure sediment flux. We find that either modelled flux method has a good linear relationship to measured flux (Figure 3; $R^2$ values of 0.98 and 0.93 for height and wavelength derived relationships respectively). While these relationships both use ripple celerity and benefit from its faster response rate, the stronger $R^2$ value for the height relationship also infers that ripple height is a better indicator of sediment flux than wavelength.

The orientations of ripple crests were slower to respond to changes in wind direction than either height or celerity. The average response time for reorientation of a mature ripple crest to a change in wind direction was 6 minutes, depending on the magnitude of both the change in wind direction and the wind speed (Supplementary Figure S2). Crestlines typically began to reorientate when the wind direction changed by more than 20°, in agreement with the observations of Sharp (1963).

These observed time lags have resonance with concepts of bedform turnover time seen in dune-scale patterns, where defect interactions are observed to be a key driver of pattern coarsening (e.g. Werner and Kocurek, 1999, Ewing et al., 2006; Day and Kocurek, 2018; Marvin et al., 2023; Marvin et al., 2025). From this perspective, although ripple celerity is driven predominantly by wind speed, the changes in wavelength and orientation generated by adjustments in erosion and deposition, and defect interactions, mean that wavelengths, orientation and potentially height should adjust at a slower rate than celerity. While we recognise that in remote planetary locations and in the sedimentary record it is not possible to measure celerity, exploration of aeolian processes in planetary environments would benefit from high resolution temporal measurements in future missions to account for lags in ripple wavelength adjustments.

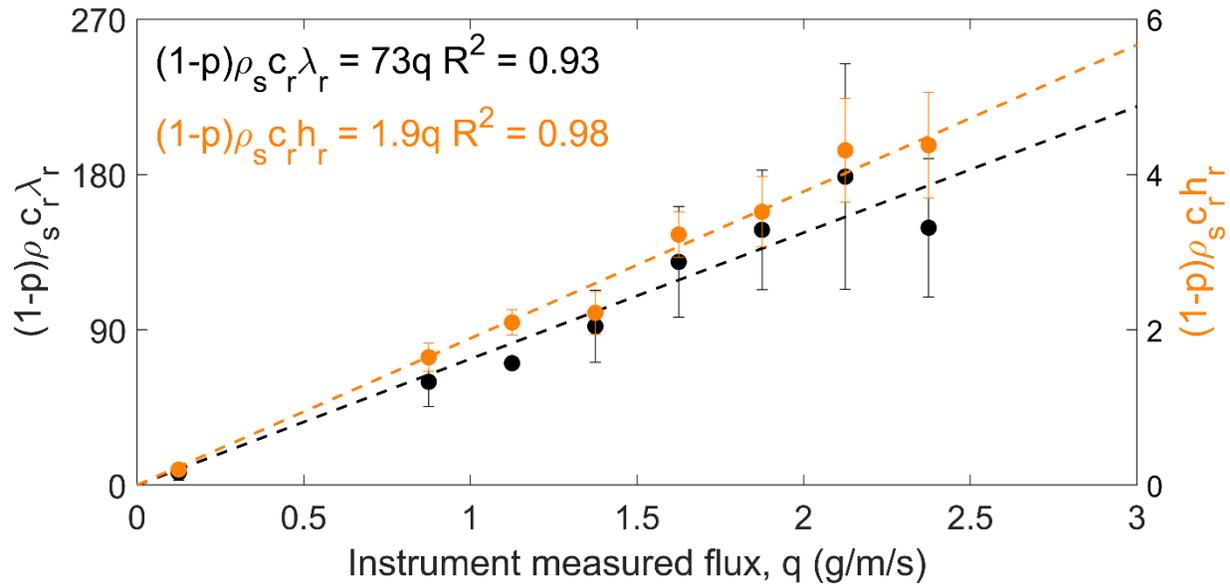

Figure 3: Comparison of wavelength-modelled and ripple height-modelled sediment flux to total field-measured flux, where p is porosity. Error bars indicate standard deviation within each measured flux bin.

Table 3  Ripple adjustment times for each dataset, identified by strongest positive cross-covariance peak with uncertainty specified with ± based on cross-covariance peak half width.

| Site | $u_*$ | Adjustment time between $u_*$ and ripple attributes (minutes) | | | | |
|---|---|---|---|---|---|---|
| | | $c_r \times h_r$ | $c_r \times l_r$ | $c_r$ | $h_r$ | $\lambda_r$ |
| Huab barchan 2014 | 0.290 ± 0.013 | 0 +1.2 | 0 + 1.3 | 0 + 1.3 | 5 ± 2.5 | - |
| Huab dome 2016 part A | 0.298 ± 0.010 | 7 ± 1.7 | 7 ± 1.6 | 7 ± 1.8 | 2 ± 0.9 | 6 ± 0.3 |
| Huab dome 2016 part B | 0.333 ± 0.016 | 0 + 1.2 | 0 + 1.0 | 0 + 1.3 | 9 ± 1.2 | 22 ± 0.3 |
| Huab dome 2018 | 0.264 ± 0.011 | 2 ± 2.1 | 1 ± 2.8 | 1 ± 2.1 | 8 ± 5.2 | 10 ± 5.5 |
| GSD 30th March 2023 | 0.541 ± 0.06 | - | - | - | 4 ± 0.3 | 4 ± 0.7 |
| GSD 30th March 2023 part B | 0.536 ± 0.068 | 1 ± 0.7 | 1 ± 0.8 | 0 + 0.6 | 7 ± 0.9 | 4 ± 1.0 |
| GSD 3rd April 2023 | 0.533 ± 0.067 | 5 ± 0.3 | 5 ± 0.3 | 6 ± 1.3 | 4 ± 0.7 | 4 ± 0.5 |
| Mean Values | | 2.5 ± 1.1 | 2.3 ± 1.2 | 2.3 ± 1.3 | 5.6 ± 1.7 | 8.3 ± 1.4 |

3.3. Saltation height

We find that aerodynamic roughness increases over a sandy, rippled surface with an increase in shear velocity (Figure 4a), in agreement with previous research (Field and Pelletier, 2018; Owen, 1964; Raupach, 1991; Sherman and Farrell, 2008). While both the Bagnold roughness equation (Eq. 1) and Charnock equation (Eq. 2) fit our data well ($R^2$ values of 0.9 and 0.89 respectively). We find that the best fit for the Charnock equation estimates a Charnock constant of 0.0155, which is closer to the wind tunnel value of 9.9 x $10^{-3}$ of Sherman et al. (2019a) rather than the Sherman et al. (2019a) field value of 0.085. This disparity may explain some of the mismatch between wind tunnel and field ripple relationships identified by Sherman et al. (2019a). Whilst it is recognised that there is a difference in how Law-of-the-Wall and Reynolds stress methods resolve $u_*$ (Lee and Baas, 2015), future studies of ripple dynamics should aim to undertake independent measurements of aerodynamic roughness and shear velocity, rather than using a modelled aerodynamic roughness value. Using our relationship between $z_o$ and shear velocity, we can identify the value of aerodynamic roughness at which the transition to the collision regime begins ($u_*/u_{*t}$ = 2.5) to be approximately 0.365 mm, indicated by the green line in Figure 4.

Our data show that there is a small increase in saltation height with increased shear velocity (Figure 4c; $R^2$=0.94). Previous research has found that saltation height is invariant with changing shear velocity (as determined by profile measurements of sediment flux and low to moderate winds, Martin and Kok, 2017). However, high resolution TLS measurements of the full saltation height profile have shown small increases in maximum saltation height with increased shear velocity (Delorme et al., 2023), particularly during strong winds (Cohn et al., 2022) where saltation height is expected to increase due to collision theory (Ralaiarisoa et al. 2020).

We find aerodynamic roughness increases with greater saltation height (Figure 4d), while changes in ripple height have a constant relationship with aerodynamic roughness. This confirms that saltation roughness drives the increase in aerodynamic roughness, as would be expected in a rough regime (Field and Pelletier, 2018). However, around a ripple height of ca. 0.0055 m, which corresponds to a $z_o$ value close to the modelled value of 0.365 mm when $u_*/u_{*t}$ = 2.5 (Figure 4b green line), we see a change in behaviour, with larger values of aerodynamic roughness.  Above this value of ripple height, $z_o$ appears to match the value derived empirically for a surface of similar physical roughness in the absence of saltation (Nield et al., 2013).  However, the standard deviation of saltation height over these larger ripples is also much greater (values greater than the green line, Figure 4d), demonstrating that these ripples may be submerged temporarily under transient sand streamers that would variably increase the saltation height and collision potential.  While more detailed studies are needed, these findings could also point to the movement of the system towards the collisional regime.

In stronger winds, it is also likely that saltation hop lengths will increase (Kok et al., 2012; Kok and Renno, 2009) and the relationship between increased hop length and ripple wavelength might not be in equilibrium, due to the longer lag time response of wavelength as compared to height (Table 3). Our findings indicate that, in addition to a horizontal length scale, the vertical length scale of ripples and its relationship to saltation cloud height must be considered. This is needed in order to better parameterise inferred flow and flux relationships for aeolian ripples, particularly when flow, and therefore transport conditions, are changing temporally.  The present results are the first that offer a way to disentangle the simultaneous influence of surface roughness (i.e. ripples) and saltation roughness (saltation height) on values of aerodynamic roughness ($z_0$) (Figure 4a-d).

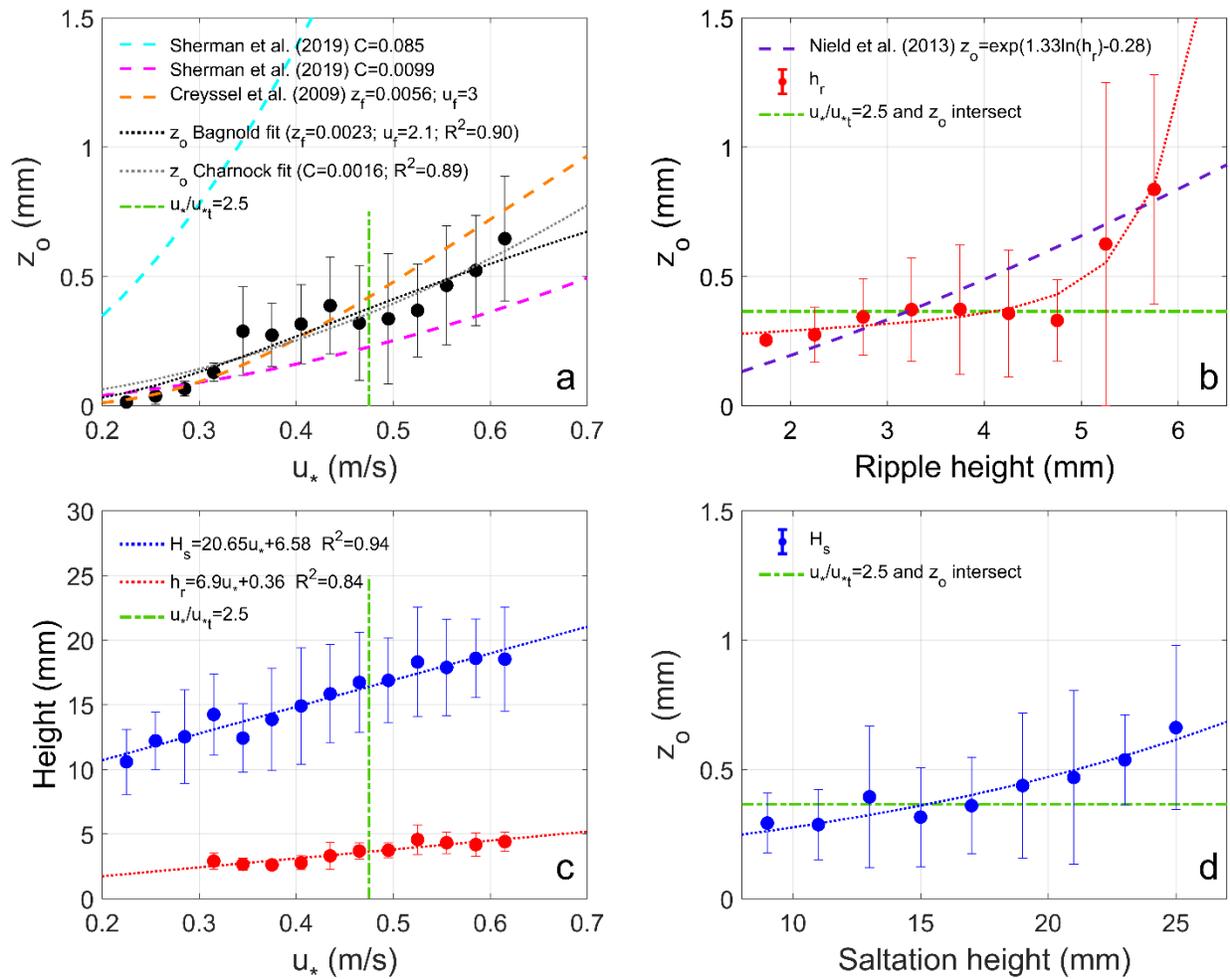

Figure 4: Variation of mean (a) aerodynamic roughness and c) saltation and ripple height with changing shear velocity. Relationship between aerodynamic roughness and b) ripple height and d) saltation height. Error bars indicate standard deviation within each shear velocity, saltation or ripple height bin. Green line identifies transition to the collision regime.

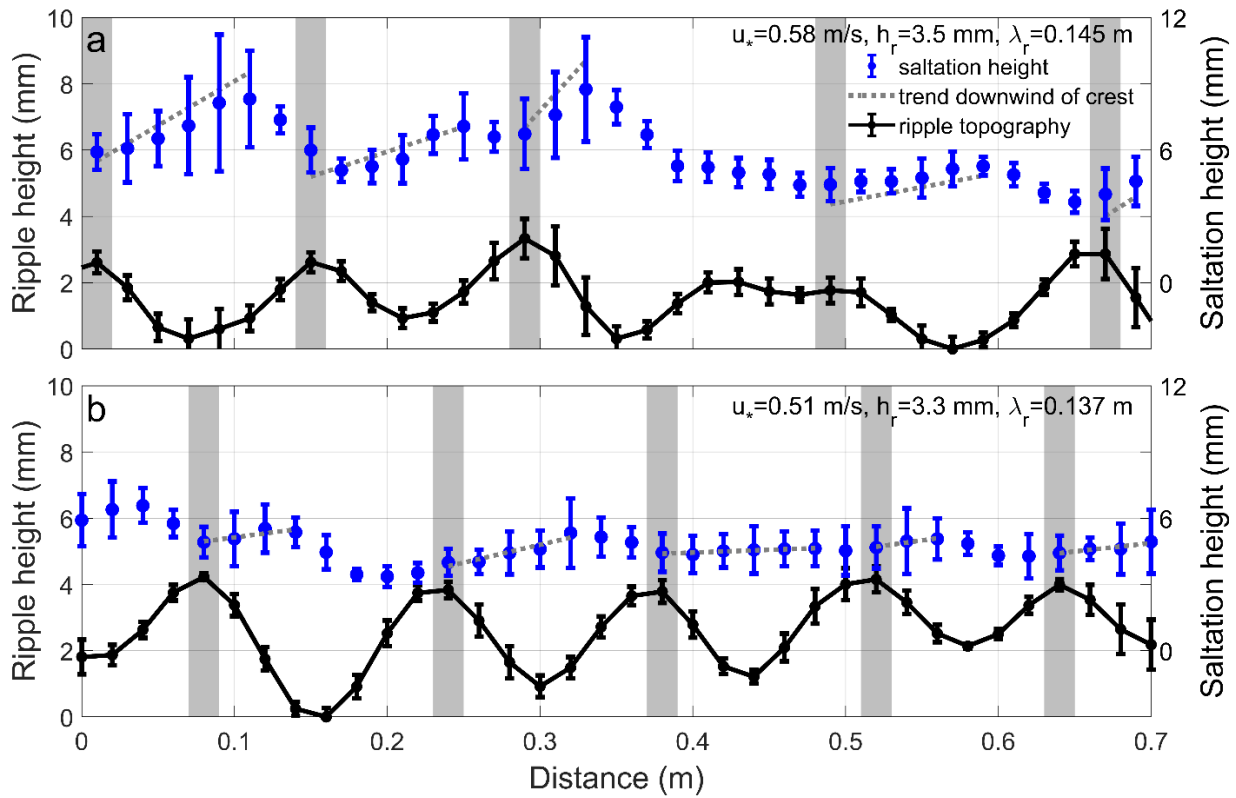

Figure 5: Examples of instantaneous saltation height over ripple topography on 30th March 2023, Great Sand Dunes National Park and Preserve. Shaded areas indicate the ripple crests. Peaks in saltation height occur 0.06 m downwind of crest, or 0.43 $l_r$. with ripple height normalised by a cross covariance of 2.52 and 0.96 for a and b respectively.

We identify field evidence for a difference in saltation dynamics over the stoss and lee slopes of ripples (Figure 5; Figure S3), in agreement with model (Anderson, 1987; Duran et al., 2014; Lester et al., 2025) and wind tunnel (Gordon and McKenna Neuman, 2011; Kelley, 2023; Kelley et al., 2025) results. On the stoss slope, we find generally smaller saltation heights relative to saltation heights on lee slopes, particularly in stronger winds (Figure S3), indicative of a greater splash density and more grain collision and ejection events over the stoss slopes (Anderson, 1987; Lester et al., 2025; Prigozhin, 1999). On the lee side, we find mean saltation heights are larger, likely because more grains bypass the lee slope (Allen, 1968; Duran et al., 2014). In the majority of cases, the initial increase in mean saltation height occurs at, or immediately downwind, of the ripple crest, with the peak in saltation height occurring approximately 0.43 $l_r$ downwind of the ripple crest (Figure 5). This peak in saltation height over ripple troughs is more robust under stronger winds (cross covariance of 2.52 $h_r$ and 0.96 $h_r$ for $u_*$ values of 0.58 and 0.51 m/s respectively; Figure 5), as more grains are transported over the ripple topography and potentially lifted into saltation (Pähtz and Tholen, 2021). We also find that the standard deviation of saltation heights over the lee slope is greater (mean value of $5.1 \times 10^{-4}$ m vs. $4.5 \times 10^{-4}$ m; Figure 4c-d), indicative of a greater variation in fall trajectories over these more sheltered slopes. This increase in the variability of saltation height over ripples during stronger winds (>2.5 $u_*/u_{*t}$) may contribute to the greater variability of aerodynamic roughness over larger ripples discussed above (Figure 4b). Our results should help to parameterise models such as those of Lester et al. (2025) and Duran et al. (2014) where the modulation of saltation over ripples is important to quantify.

Future research should examine the dynamics of ripples over a greater range of grain sizes and wind regimes to extend our understanding of these self-organising patterns (Anderson, 1990; Baas, 2002; Coco and Murray, 2007; Landry and Werner, 1994). Further, whilst our findings elucidate the key role of surface features and saltation trajectories in aeolian processes, more studies will help quantify how form-flow-saltation dynamics change under strong winds.

4. Conclusions

We find that ripple dynamics are good indicators of shear velocity and sediment flux, with ripple celerity possessing a stronger relationship with flow conditions than either ripple height or wavelength. Although ripple wavelength and height can also be used to infer flow conditions, ripple wavelength is slower to respond to changes in wind speed, particularly when wind speeds are decreasing, and so caution should be used when utilizing ripple wavelength to infer instantaneous flow conditions. This is significant because wavelength dimensions are often the only available data from contemporary planetary landscapes, or derived from supercritically climbing aeolian ripples (sensu Hunter, 1977) within the rock record. In these instances, we suggest wavelengths might not represent the full range of fluctuating wind conditions, but rather approximate previous mean sediment transport conditions that were sustained for a long enough period of time that the ripples had fully adjusted. While ripple height and saltation height both increase with shear velocity, the impact of shear velocity on aerodynamic roughness – imparted by grains, sediment transport and form effects – is more nuanced. Critically, we also find a non-linear relationship between ripple celerity and shear velocity at high wind speeds. While a transition to a collision regime may in part help to explain both this non-linear celerity relationship and the switch from saltation height to ripple height as the key driver of aerodynamic roughness, more experiments are needed to test these hypotheses that capture coincident sediment transport and surface morphology for different grain sizes and wind conditions.

## Acknowledgements

Funding is gratefully acknowledged from the National Geographic Society (GEFNE110-14 and CP-029R-17), British Society for Geomorphology, Natural Environment Research Council, UK and National Science Foundation, USA NSFGEO-NERC (NE/R010196/1; NSF #1829513; NSF#1829541; GEF-1025). The authors thank J. Mayaud, M. David, S. Muinjo and B. Shiyanga for field assistance. We are grateful to the SoGES laboratory team who undertook the grain size analysis. We also acknowledge I. Matheus, J. Kazeurua, L. Uahengo, the Namibia Ministry of Environment and Tourism, NCRST and Skeleton Coast National Park Rangers for advice and site access (permits 1913/2014; 2051/2015; 2168/2016, RPIV00022018). F. Bunch was instrumental in the GSD fieldwork (permit GRSA-2018-SCI-004). J.M. Nield gratefully acknowledges the Department of Geology and Geophysics, Texas A&M University, where she was supported by a Michel T. Halbouty Visiting Chair during processing some of the data for this manuscript. Data processing used IRIDIS Southampton Computing Facility. We thank Joel Sankey, Mathieu Lapôtre and two anonymous reviewers for very useful comments that helped improve this manuscript.

Conflict of Interest

The authors declare no conflicts of interest relevant to this study.

Open Research

All raw data and processing scripts (Nield et al., 2025a; Nield et al., 2023a; Nield et al., 2023b) along with processed ripple dynamics (Nield et al., 2025b) are available on the NERC EDS National Geoscience Data Centre, UK.

Author Contributions

Funding acquisition and conceptualisation: JMN, GFSW, MCB, JB, KTC, MHTH, PC, AV; Data curation: JMN; Formal analysis: JMN; Methodology and data collection: JMN, GFSW, MCB, PD, PC, AV, MHTH, DPG, NSW; Writing (original draft/review and editing): all

author accepted manuscript
Nield et al. (2025) Journal of Geophysical Research: Earth Surface, 130, e2025JF008616
http://dx.doi.org/10.1029/2025JF008616 - available open access from publisher

author accepted manuscript
Nield et al. (2025) Journal of Geophysical Research: Earth Surface, 130, e2025JF008616
http://dx.doi.org/10.1029/2025JF008616 - available open access from publisherLorenz, R. D. (2011), Observations of wind ripple migration on an Egyptian seif dune using an inexpensive digital timelapse camera, *Aeolian Research*, *3*(2), 229-234, doi:10.1016/j.aeolia.2011.01.004.

Lorenz, R. D., and A. Valdez (2011), Variable wind ripple migration at Great Sand Dunes National Park and Preserve, observed by timelapse imaging, *Geomorphology*, *133*(1-2), 1-10, doi:10.1016/j.geomorph.2011.06.003.

Marticorena, B., and G. Bergametti (1995), Modeling the atmospheric dust cycle: 1. Design of a soil-derived dust emission scheme, *J. Geophys. Res*, *100*(D8), 16415-16430, doi:10.1029/95JD00690.

Martin, R. L., and D. J. Jerolmack (2013), Origin of hysteresis in bed form response to unsteady flows, *Water Resources Research*, *49*(3), 1314-1333, doi:10.1002/wrcr.20093.

Martin, R. L., and J. F. Kok (2017), Wind-invariant saltation heights imply linear scaling of aeolian saltation flux with shear stress, *Science Advances*, *3*(6), doi:10.1126/sciadv.1602569.

Martin, R. L., and J. F. Kok (2018), Distinct thresholds for the initiation and cessation of aeolian saltation from field measurements, *Journal of Geophysical Research: Earth Surface*, *0*(ja), doi:10.1029/2017JF004416.

Marvin, M. C., M. G. A. Lapôtre, A. Gunn, M. Day, and A. Soto (2023), Dune interactions record changes in boundary conditions, *Geology*, 51(10), 947-951, doi:10.1130/g51264.1.

Marvin, M. C., M. G. A. Lapôtre, J. Radebaugh, and W. Bo (2025), From Xanadu around and back: A ca. 11,000 km Journey of windblown sand revealed by global dune patterns on Titan, *Geophysical Research Letters*, 52(5), e2024GL112760, doi:https://doi.org/10.1029/2024GL112760.

Mayaud, J. R., R. M. Bailey, G. F. S. Wiggs, and C. M. Weaver (2017), Modelling aeolian sand transport using a dynamic mass balancing approach, *Geomorphology*, *280*, 108-121.

Nield, J. M. (2011), Surface moisture induced feedback in aeolian environments, *Geology*, *39*(10), 915-918, doi:10.1130/G32151.1.

Nield, J. M., M. C. Baddock, and G. F. S. Wiggs (2025a), Surface and meteorological data of ripples, saltation and protodune dynamics at Medano Creek, Great Sand Dunes National Park and Preserve, Colorado, USA in March and April 2023, (Dataset), NERC EDS National Geoscience Data Centre, doi:10.5285/4b86b7a5-51ee-47b8-836d-a72cfd2be26b.

Nield, J. M., M. C. Baddock, G. F. S. Wiggs, P. Delorme, A. Valdez, N. Wallum, and D. Goss (2023a), Surface and meteorological data of ripple and saltation dynamics at Medano Creek area, Great Sand

**Quantifying form-flow-saltation dynamics of aeolian sand ripples**

Joanna M. Nield[1]*, Matthew C. Baddock[2], Giles F.S. Wiggs[3], Jim Best[4], Kenneth T. Christensen[5], Pauline Delorme[6], Andrew Valdez[7], Nathaniel R. Bristow[8], Martin H.T. Hipondoka[9], Daniel P. Goss[1], Natasha S. Wallum[3], Philippe Claudin[10]

1 School of Geography and Environmental Science, University of Southampton, Southampton, SO17 1BJ, UK

2 Geography and Environment, Loughborough University, Loughborough, LE11 3TU, UK

3 School of Geography and the Environment, University of Oxford, Oxford, OX1 3QY, UK

4 Departments of Earth Science and Environmental Change, Geography & GIS, and Mechanical Science & Engineering, and Ven Te Chow Hydrosystems Laboratory, University of Illinois at Urbana-Champaign, Urbana, IL 61801, USA

5 Department of Mechanical Engineering, University of Colorado Denver, Denver, CO, 80204, USA

6 Laboratoire de Géologie, Ecole Normale Supérieure, CNRS, PSL Research University, Paris, France

7 Great Sand Dunes National Park and Preserve, National Park Service, Alamosa, CO, USA

8 Astrin Biosciences, Inc., Saint Paul, MN, 55114, USA

9 Department of Environmental Science, University of Namibia, Private Bag 13301, Windhoek, Namibia

10 Physique et Mécanique des Milieux Hétérogènes, CNRS, ESPCI Paris, PSL Research University, Université Paris Cité, Sorbonne Université, Paris, France

**Contents of this file**

> Text S1
>
> Figures S1 to S3
>
> Table S1

**Additional Supporting Information (Files uploaded separately)**

> Caption for Movie S1

**Introduction**

The supporting information includes additional methodological details about how the data were processed (Text S1) and additional figures and tables that relate to this processing and give context to the article.

**Text S1: Supporting Information Data Processing Methods**

In 2022, the surface was measured simultaneously using two adjacent Leica TLS (P20 and P50) while in 2023 a single P50 was used. Each TLS measured an approximately 1 x 1 m square area, approximately 9 m from the TLS head that was positioned at a height of around 2 m and measured parallel to the wind, so as to reduce any distortion due to ripple migration during each scan and minimise obscuration of the surface by saltation streamers. The horizontal scan resolution was either 0.8 or 1.6 mm at 10 m that took 3 or 1.5 minutes respectively to complete the scanned square. These settings were varied depending on wind speed to ensure that the ripples did not migrate further than one wavelength between scans. The TLS vertical resolution has been calculated at $5.5 \times 10^{-4}$ m (Baddock et al., 2018). Wind speed and direction were measured downwind of each rippled square at 10 Hz using a Campbell Scientific CSAT3 3D sonic anemometer at 0.24 m above the surface. The Reynolds decomposition approach was used to convert the wind speed vectors into a shear velocity and to determine wind angle variability for each scan time period (Baddock et al., 2011; Mayaud et al., 2017; Weaver and Wiggs, 2011). Saltation flux measurements were recorded immediately downwind of each scanned square using i) a Sensit that was positioned so that the sensor base was flush with the surface and the sensor top extended to a height of 0.014 m above the surface, and ii) Wenglor optical gate sensors at heights of 0.02 m and 0.05 m. Sensit particle impact count data, n, were converted to a saltation point flux, $q_s$, following the methods of Delorme et al. (2023) for each time period, using a grain volume (V), assuming spherical grains and a measured saltating grain size, $d_{50}$, of 288 µm, a sediment density, $\rho_s$, of 2650 kg m$^{-3}$, and sensor width (W) of 0.024 m,

$$q_s = \frac{nV\rho_s}{W} \qquad (3)$$

Wenglor particle count data (n) was converted to a saltation point flux using Equation 3, where W was equal to 0.03 m. Shear velocity threshold ($u_{*t}$) was calculated by fitting sediment flux (q) to Equation 4 (Ungar and Haff, 1987), where c is a fitting parameter, $u_*$ is shear velocity,

$$q = c\frac{u_*^2 - u_{*t}^2}{u_{*t}^2} \qquad (4)$$

TLS data were post-processed following similar methods to Martin and Jerolmack (2013) and Nield (2011) by first applying a 0.1 m radial filter with an angle of 35° to separate out points above the surface (saltons) from surface returns (Nield and Wiggs, 2011). The surface points were gridded at a cell width of 0.002 m and further cleaned using a mean moving window filter of 0.05 x 0.05 m. This produced a continuous rippled surface (Supplementary Figure 1b). Although the measurements were undertaken on a flat, sandy surface, to remove any larger scale surface gradients associated with the nearby stream channel (c. 0.01 in some instances), each dataset was detrended using the overall mean surface slope for the total experiment duration.

Points that were classified within the above-surface subset by the radial filter were gridded at the same 0.002 m cell width as the surface points, and the distance between the maximum non-surface point and the surface grid was used to determine the maximum saltation height at each cell (Nield and Wiggs, 2011).  This high-resolution dataset was used to determine the maximum saltation height, $H_s$, over the stoss and lee sides of individual ripples.  For high resolution measurements of saltation heights over individual ripples, three consecutive transects were translated according to the measured migration rate and the average ripple morphology and saltation heights calculated with a 0.02 m resolution.  For bulk measurements of slope-saltation relationships, grid points were classified as either stoss or lee slopes based on the direction of the smoothed ripple topography.  The saltation heights were then grouped by which slope they were measured over and the mean of these 0.002 m values were binned separately with changing shear velocity.  To remove the local topographic effect of the ripples themselves, more general relationships between saltation height, aerodynamic roughness, ripple height and shear velocity were calculated over a coarser grid (0.2 x 0.2 m).  This grid size was chosen to be the same length as a typical ripple wavelength and ensured that the ripple trough was the minimum surface point for the maximum saltation height calculation.

Surface flux was calculated as erosion and deposition rates by differencing DEMs of each scan.  Total flux was calculated by integrating the curve of the point-based flux measurements from the Sensit and Wenglor samplers, along with the surface flux from the ripple migration measurements. Quadratic relationships for the point-based and total flux were fitted to Equation 4 following methods of Delorme et al. (2023) to determine $u_{*t}$ with a mean value from the different transport datasets of 0.19 m/s (Supplementary Figure S1).

TLS measurements in the Huab Valley followed the methods used for the main Great Sand Dunes dataset with a single Leica P20 TLS.  There were no sediment flux measurements and wind velocity was measured by a 3D sonic at 0.5 m height or by a single cup anemometer at a height of 0.24 m logging every 10 seconds.  We converted mean velocity measured with the cup anemometer to shear velocity assuming an aerodynamic roughness value of 1.9 x $10^{-5}$ m (Supplementary Table S1) and the Law-of-the-Wall equation.  The $d_{50}$ grain size of the sand surface on the dome and barchan dunes were 402 μm and 341 μm respectively.

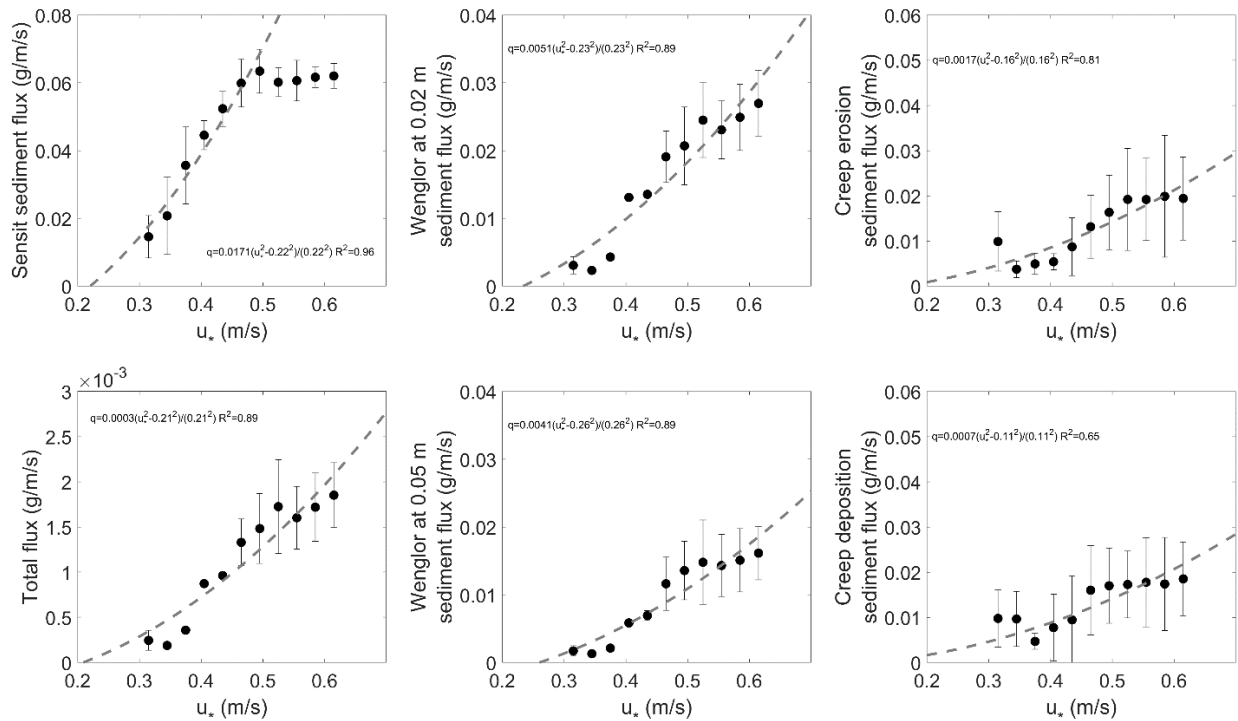

**Figure S1**: Shear velocity threshold for sand entrainment calculated from different flux measurements (mean $u_{*t}$ value of 0.19 m/s for total flux and erosion creep flux). The differences in threshold highlight the vertical complexity of saltation and bed deformation with an increase in wind speed being required to initiate the movement of the ripple surface vs. more sporadic saltation around the threshold windspeed. These threshold differences are also indicative of saltation hysteresis (Martin and Kok, 2018) along with the difficulties in defining what is meant by a threshold with respect to transport or morphological change.

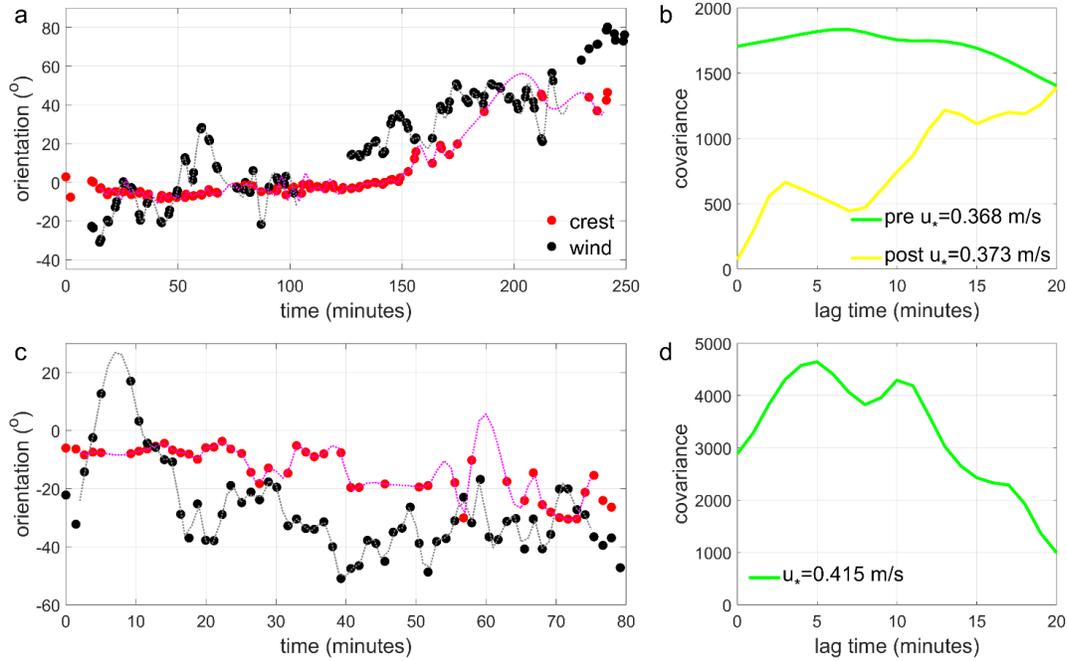

**Figure S2**: Crestline orientation and wind direction for ripples on (a) 5th April 2022 and (c) 4th April 2023. Cross covariance and lag times between these orientations in (b) 2022 and (d) 2023. The 2022 data is split into two section pre and post 100 minutes due to a gap in wind measurements.

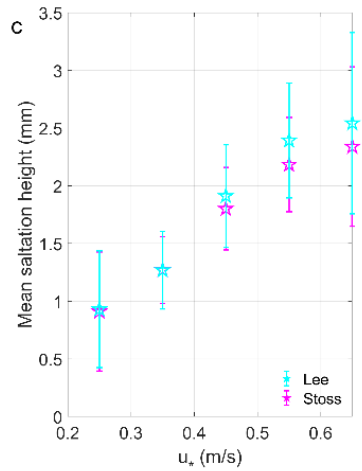

**Figure S3**: Mean saltation height for all GSD experiments over stoss and lee sides of ripples. Error bars indicate standard deviation within each shear velocity bin.

**Table S1**: Aerodynamic roughness calculations for similar surfaces in the same location where sediment transport was occurring (Skeleton Coast) Average of $1.9 \times 10^{-5}$ m is similar to measurements close by on a sand sheet by Weaver (2008).

| Location | Height of CSAT sonic anemometer (m) | $z_o$ (m) | Transport |
|---|---|---|---|
| Centre area of 0.12 m high protodune | 0.1 | 2.7E-05 | constant |
| centre area of barchan | 0.5 | 3.1E-05 | constant |
| side of barchan | 0.5 | 1.9E-05 | constant |
| side of barchan | 0.5 | 1.2E-05 | constant |
| brink of barchan | 0.5 | 7.0E-06 | constant |
| Sand sheet (Weaver, 2008) | 0.3 | 1.9E-05 | constant |

**Movie S1**: Ripple DEMs through time on the 30th March 2023. Wind direction left to right. Surface topography (left) and identified crests (right). Crests used for orientation analysis in white. Central cross section changes through time (bottom left). File name: Movie_S1_Nieldetal_2025